\documentclass[useAMS]{mn2e}

\usepackage{ulem} 
\usepackage{graphicx}
\usepackage{amsmath}
\usepackage{ulem}
\usepackage{rotating}
\usepackage{amssymb}

\def\lapprox{\hbox{\lower .8ex\hbox{$\,\buildrel < \over\sim\,$}}}
\def\gapprox{\hbox{\lower .8ex\hbox{$\,\buildrel > \over\sim\,$}}}

\title[COMET 2I/BORISOV]{
Interstellar Comet 2I/Borisov exhibits a structure similar to native Solar System comets\thanks{    
Based on observations with the NASA/ESA {\it Hubble
Space Telescope}, obtained at the Space Telescope Science Institute,
which is operated by AURA, Inc., under NASA contract NAS 5-26555.
}
}
\author[F.\,Manzini et al.]{
  F.\,Manzini$^{1}$\thanks{E-mail: manzini.ff@aruba.it}, V.\, Oldani$^{1}$,
  P.\,Ochner$^{2,3}$, and L.\,R.\,Bedin$^{2}$\\
$^{1}$Stazione Astronomica di Sozzago, Cascina Guascona, I-28060 Sozzago (Novara), Italy\\
$^{2}$INAF-Osservatorio Astronomico di Padova, Vicolo dell'Osservatorio 5, I-35122 Padova, Italy\\
$^{3}$Department of Physics and Astronomy-University of Padova, Via F. Marzolo 8, I-35131 Padova, Italy
}

\begin{document} 

\date{\textit{Letter:} Accepted 2020 April 1. Received 2020 April 1; in original form 2019 November 22.}

\pagerange{\pageref{firstpage}--\pageref{lastpage}} \pubyear{201X}

\maketitle
 
\label{firstpage}

\begin{abstract}
We  processed images  taken  with the  \textit{Hubble Space  Telescope
  (HST)} to investigate  any morphological features in  the inner coma
suggestive of a  peculiar activity on the nucleus  of the interstellar
comet 2I/Borisov.
The  coma shows  an evident  elongation,  in the  position angle  (PA)
$\sim$0$^\circ$-180$^\circ$  direction, which  appears related  to the
presence  of a  jet originating  from a  single active  source on  the
nucleus.   A   counterpart   of   this   jet   directed   towards   PA
$\sim$10$^\circ$ was detected  through analysis of the  changes of the
inner coma morphology on \textit{HST}  images taken in different dates
and processed  with different  filters.  These findings  indicate that
the nucleus is  probably rotating with a spin axis  projected near the
plane of the sky and oriented at PA $\sim$100$^\circ$-280$^\circ$, and
that   the    active   source   is   lying    in   a   near-equatorial
position.  Subsequent  observations  of  \textit{HST}  allowed  us  to
determine   the  direction   of  the   spin  axis   at  RA   =  17h20m
$\pm$15$^\circ$ and Dec = $-$35$^\circ$ $\pm$10$^\circ$.
Photometry of  the nucleus on  \textit{HST} images of 12  October 2019
only span $\sim$7 hours, insufficient to reveal a rotational period.
The morphology exhibited by the  interstellar comet 2I/Borisov is very
similar to  that of  comet C/2014\,B1  suggesting that  the activation
processes are the same observed in the Solar System native comets.

\end{abstract}


%
\section{Introduction}
\label{introduction}
%
%
Comet 2I/Borisov is only the  second interstellar object known to have
passed  through  the  solar  system  (Guzik  et  al.\,  2019a,  Ye  et
al.\,2019);  it  therefore  provides   an  invaluable  opportunity  to
investigate the  structure and  dust characteristics of  small objects
presumably formed in another, distant star system.

This comet  was observed in images  taken on 10 and  13 September 2019
with the WHT and GNT telescopes,  when it showed only an extended coma
and a faint,  broad tail. The color of the  comet was slightly reddish
with  a  $(g^\prime-r^\prime)$   color  index  of  0.66$\pm$0.01\,mag,
compatible with solar system comets.  Its observed morphology was best
explained by  dust with a low  ejection speed (44$\pm$14\,m\,s$^{-1}$)
for  $\beta=1$ particles,  where $\beta$  is  the ratio  of the  solar
gravitational  attraction to  the solar  radiation pressure  (Guzik et
al.,  2019b).   Almost  simultaneously   the  comet  was  observed  by
D.\,Jewitt and J.\,Luu with the NOT  telescope on six dates between 13
September and 4 October, and  the early results prompted D.\,Jewitt et
al. to submit a successful DDT  proposal to observe comet Borisov with
the \textit{Hubble  Space Telescope}.  Some results were  published in
Jewitt et al.\ (2019a).

A total  of 24 images of  the comet, of  260s each, were taken  by the
\textit{HST} WFC3 camera in the  F350LP broad-band filter (centered at
584.94 nm)  on 12 October  2019 in  4 consecutive orbits.   After they
were released to  the public we could use them  to perform an analysis
of the morphology of the coma of the comet, which is the focus of this
Letter.   Additional  analyses were  performed  after  release of  new
images taken by \textit{HST} during  single orbits on three successive
dates: 16 November and 9 December 2019, and 3 January 2020.
%
%
\section{Methods} 
%
%
In order to perform a morphological  analysis of the inner coma, where
structures due to the emission of  gas and dust from active sources on
the  nucleus are  more  easily detectable,  we  co-added and  averaged
separately  the  four   series  of  six  \textit{HST}   images  of  12
October.  The comparative  analysis of  the four  series did  not show
obvious structural  variations over  the $\sim$7\,h  imaging time-span
(from 13h44m39s UT  to 20h42m23s UT of 12 October  2019), therefore we
decided to stack the 4 series of \textit{HST} images in a single image
corresponding to a total exposure of 6240\,s (104\,min), to obtain the
highest possible signal to noise ratio.

To highlight  the structures of  the inner coma, we  applied different
algorithms, often used for this  type of analysis: median subtraction,
division by $1/r$, division by $1/r^{1.2}$ (Samarasinha et al.\,2014),
Larson-Sekanina  with 30$^{\circ}$  angle (Larson  et al.\,1984).  The
comparison of the results of the  four methods showed that the details
highlighted were always the same in the same place, however the latter
process provided the highest signal to noise ratio (Fig.\,\ref{F1}).

We  also applied  false  color palette  or  isophote visualization  to
enhance  the   visibility  of  the   details,  as  well  as   a  polar
transformation, centered  on the optocenter, to  the processed images,
with  the aim  of highlighting  and determining  the precise  position
angle (PA)  of any structures  revealed by the  processing algorithms.
Knowing that the nominal resolution of the \textit{HST} WFC3 camera is
$\sim$0.03977\,arcsec/pixel   (Bellini  et   al.\  2011,   Gennaro  et
al.\ 2018), corresponding  to 81.1\,km on the plane of  the sky at the
distance  of the  comet,  it was  then easy  to  convert the  measured
distances from pixels into km.

Finally, we also performed photometric measures of the nucleus on each
of  the original  \textit{HST}  images to  determine its  instrumental
magnitude. We applied the same parameters used by Bolin et al. (2019):
a photometric circle with a  5-pixel diameter (0.2 arcsec) centered on
the comet’s optocenter  (assumed to correspond to the  nucleus) and an
annulus  with internal  and external  radii of  6 and  20 pixels  (0.8
arcsec), respectively, to measure the mean background value.

%
\section{Analysis of the Comet} 
%
\subsection{Dust Tail} 

\begin{figure}
\begin{center}
\includegraphics[width=84mm]{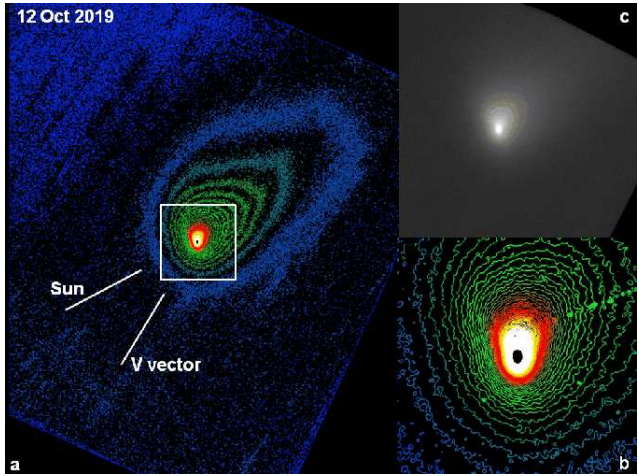}
\caption{
Stacked  \textit{HST}  image  shown   in  isophote  color  palette  to
highlight  the shape  and direction  of  the tail.  The difference  in
brightness between each isophote is 75 ADU. The farthest blue isophote
is just 20 ADU  above the sky background. The length  of the dust tail
is 40 arcsec ($8\times10^{4}$\,km) on the sky plane. (b) Magnification
of   the    area   surrounding   the   nucleus    (10   arcsec   side,
$2\times10^{4}$\,km). (c) Original \textit{HST}  image as in (a), 0.5x
resized.
\label{F1}
}
\end{center}
\end{figure}

The stacked \textit{HST} image of 12 October was processed in isophote
color palette, to  determine the shape and direction of  the tail.  We
applied  a brightness  gap between  each isophote  as low  as 75  ADU,
ranging from 19307  ADU at the peak brightness of  the comet’s nucleus
to  1820 ADU  of the  farthest blue  isophote, just  20 ADU  above the
sky-background,   in   order   to    detect   the   faintest   details
(Fig.\,\ref{F1}).  The   orientation  of   the  tail  appears   to  be
approximately  aligned  with  the   anti-solar  vector  in  the  image
visualized  with  this  method.   Recently,  Jewitt  et  al.\  (2019a)
reported that  the visible  portion of  the tail  is limited  to about
60\,arcsec in length by sky noise and field structures on images taken
with  Earth-based telescopes.  However,  in  the stacked  \textit{HST}
images, the farthest point where the end of the tail could be detected
is about 970 pixels from the  nucleus, (40 arcsec), corresponding to a
sky-plane  length  $L=8.1\times10^{4}$\,km  and,  assuming  it  is  in
antisolar   direction,   to   an   estimated   true   length   $L_{\rm
  T}=8.1\times10^{4}  \times \sin{\alpha^{-1}}  = 2.4\times10^{5}$\,km
(with  $\alpha=20^\circ$,  where  $\alpha$   is  the  phase  angle  as
calculated           with           the           \textit{JPL/Horizon}
software\footnote{\texttt{https://ssd.jpl.nasa.gov/horizons.cgi}}).

The high density of the isophotes  in the area surrounding the nucleus
(white-yellow-red isophotes,  panel\,b in  Fig.\,1) suggests  that the
tail could  partially originate from an  almost isotropic distribution
of dust, although the elongation of  the inner coma is also suggesting
a  strong  contribution  of  a  radial  emitting  structure.   On  the
contrary, the  rapid decrease of  the density of the  isophotes moving
away from the nucleus  towards the end of the tail  is indicative of a
wide  scattering of  the  emitted  dust already  at  a relatively  low
distance from the nucleus.

The length of  the tail can provide useful information  about the size
of the dust grains of which it  is made. Within the tail, the dust and
gas emitted from  the nucleus are decoupled, and  the only significant
forces affecting the grain trajectories  are the solar gravity and the
radiation  pressure.   Both  forces  depend   on  the  square  of  the
heliocentric  distance,  but  work in  opposite  directions  (Vincent,
2010).  Their sum  can be  seen as  a reduced  solar gravity,  and the
motion of dusts follows the equation
\begin{equation}
 m \times a = (1-\beta) \times ({\rm Solar\,gravity})
\label{EQ1}
\end{equation}
where $\beta$  is the ratio (radiation  pressure)/(solar gravity), and
is  inversely proportional  to the  size of  the grains  for particles
larger than 1\,$\mu$m (Bohren \& Huffman, 1983).

\begin{figure}
\begin{center}
\includegraphics[width=84mm]{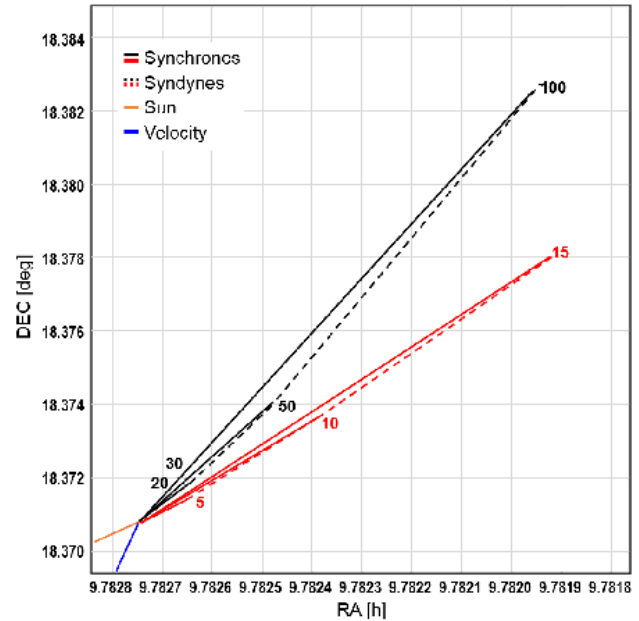}
\caption{
Two-dimensional  modeling of  the  dust  tail, made  by  means of  the
Finson-Probstein diagram.  The synchrones (black solid  lines) and the
syndynes (black  dashed lines) originated  from dust emissions  of the
nucleus occurred 15  to 120 days before 12 October  2019 are shown for
$\beta$=0.01. In red  the synchrones and syndynes from  emissions 5 to
15 days before the same date for $\beta$=0.3. For these higher $\beta$
values the model  approximates the scattering of the  dust as observed
on the \textit{HST} images.
\label{F2}
}
\end{center}
\end{figure}

We performed  a two-dimensional  simulation of the  dust tail  made by
means of the Finson-Probstein diagram (Comet Toolbox by J.B. Vincent
\footnote{\texttt{http://www.comet-toolbox.com}})   up  to   120  days
earlier    than   12    October   2019    using   different    $\beta$
(Fig.\,\ref{F2}).  Considering   the  approximations   introduced,  we
obtained a  length and shape of  the tail in the  numerical simulation
corresponding to that measured on the \textit{HST} image for values of
$\beta$ between 0.01 and 0.03; it  can therefore be estimated that the
dust grains in the distal tail have radii between 30 and 100 $\mu$m.

%
\subsection{Coma} 
\begin{figure}
\begin{center}
\includegraphics[width=80mm]{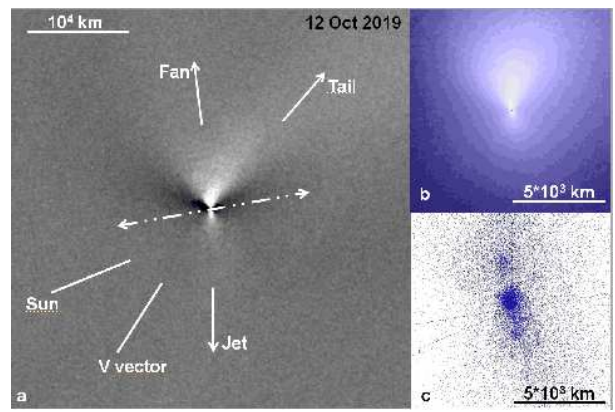}
\caption{
Panel  a:  \textit{HST}  image,  processed  with  the  Larson-Sekanina
filter, showing the jet in PA 180$^\circ$ and the supposed position of
the spin axis. (b): same  image, processed with 1/r filter, visualized
in isodensity  contours; (c) computer  model of the morphology  of the
inner coma following the emission of dust from a single active area in
equatorial position,  showing features consistent with  those observed
in the \textit{HST} images.
\label{F3}
}
\end{center}
\end{figure}

%
%
The composite \textit{HST} image of  12 October 2019 clearly shows the
comet as non- stellar and the  inner coma anisotropic and elongated in
N-S  direction ($<$3\,arcsec  or 75  pixels, Bolin  et al.\,2019),  in
contrast with  the projected direction  of the tail which  is oriented
towards  PA  300$^\circ$. This  elongation  becomes  more obvious  and
prominent in the composite image. The elongation appears to be related
to the presence  of a jet, that  we could detect with  all the spatial
filters applied. At least part of the dust coma most likely originates
from  the emission  of  this  jet, well  visible  directed towards  PA
180$^\circ$   in  the   image  processed   with  the   Larson-Sekanina
filter.  The high  resolution  of  the WFC3  camera  shows no  visible
sub-structures  of this  jet,  suggesting that  it  originates from  a
single active source on the nucleus of the comet (Fig.\,\ref{F3}).

%
%
%
\begin{figure}
\begin{center}
\includegraphics[width=80mm]{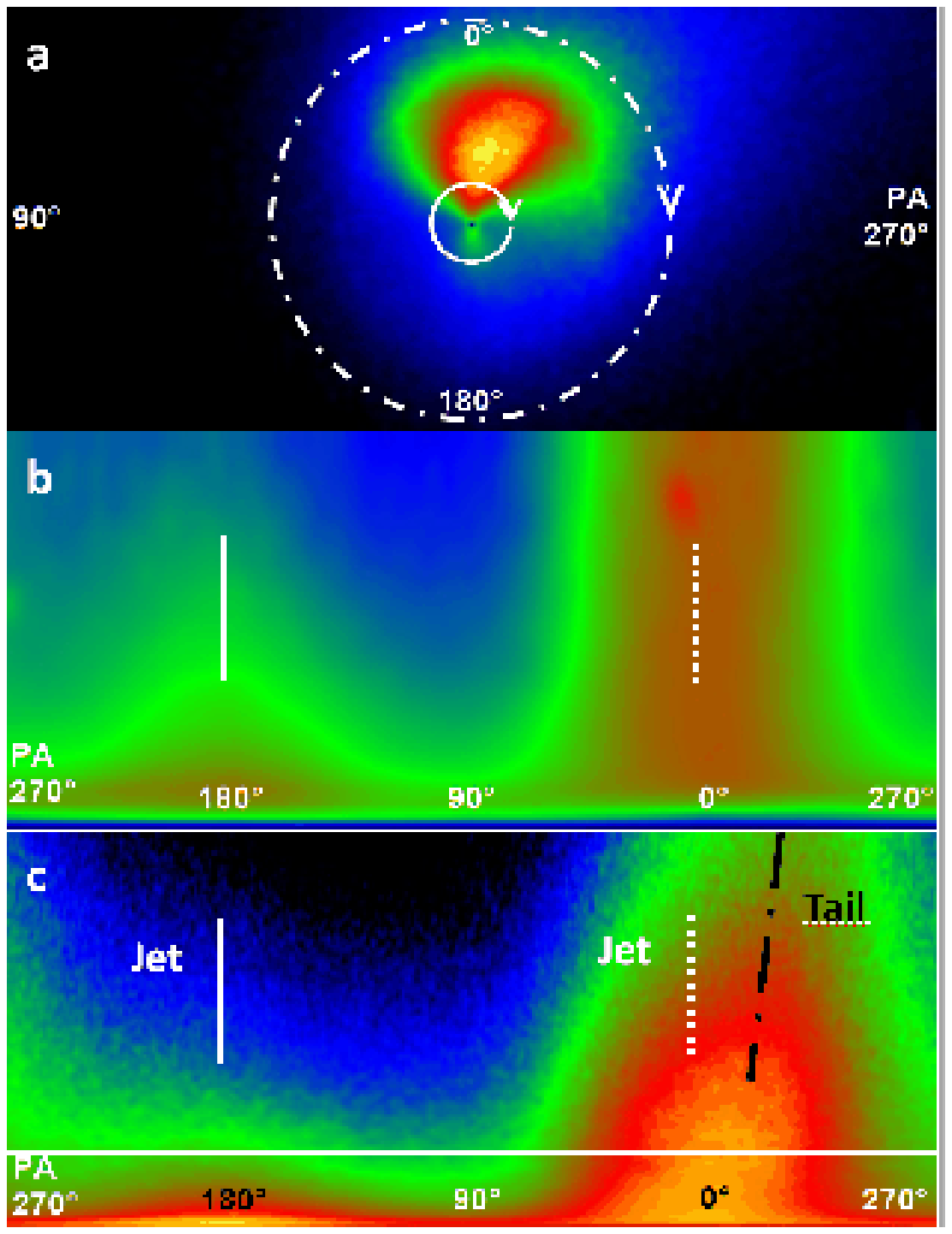}
\caption{
The original \textit{HST}  image processed with the  1/r filter, shown
in false  colors in panel  a, was  transformed in polar  projection to
identify the morphological  details in the inner coma up  to 30 pixels
($2.4\times10^{3}$\,km  -  solid white  circle)  and  their PA  (panel
b). The  brightest condensations at  PA 180$^\circ$ and  PA 10$^\circ$
correspond to the jets deriving from  the emissions of a single active
area. (c) Polar  projection up to 150  pixels ($1.2\times10^{4}$\,km -
dash and dot circle) shows the extent of the features on a wider scale
and the development of the tail at a different PA than the jet.
\label{F4}
}
\end{center}
\end{figure}
%

The emitted dust  appears scattered at a relatively  low distance from
the nucleus and  is soon deflected by the solar  radiation pressure to
merge into the  tail. This appearance is indicative of  a low ejection
speed  and  possibly related  to  abundance  of  dust particles  of  a
relatively small size.

We made  further in depth analysis  in search for other  structures in
different position  angles. Actually, we found  a fan-shaped structure
towards PA 10$^\circ$, in a contralateral position with respect to the
jet. The  two structures are  probably related, as they  appear almost
symmetrical on the  opposite sides, thus most  likely originating from
the same single active source on a rotating nucleus (Fig.\,\ref{F4}).

\subsection{Photometry of the Nucleus}
\begin{figure}
\begin{center}
\includegraphics[width=80mm]{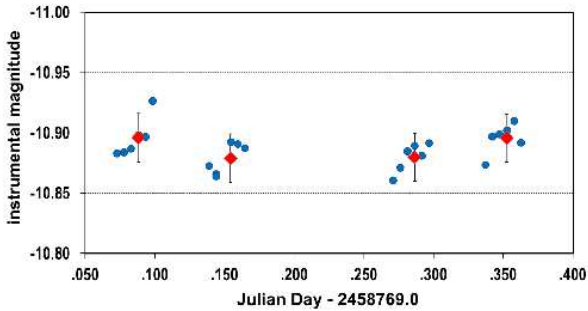}

\caption{
Photometry  with  instrumental values  on  the  nucleus of  2I/Borisov
performed on the original \textit{HST} images taken on 12 Oct 2019 and
spanning 7h. The blue dots are the measurements of each image, the red
diamonds are the calculated mean magnitude of each series of images.
\label{F5}
}
\end{center}
\end{figure}

The images  taken with  \textit{HST} are  tracked at  the differential
motion rate of the comet with respect to the stars, therefore we could
only perform a  photometry to determine the  instrumental magnitude of
the nucleus.  The use of  a small  aperture (5-pixel diameter)  of the
photometric circle  removes efficiently the  diffuse light due  to the
dust in  the coma  and enables  to detect  the smallest  variations in
brightness due  to discrete emissions  related to the rotation  of the
nucleus  (Lamy  et  al.\,  1998a, 1998b).  The  measured  instrumental
magnitudes show  small fluctuations  suggestive of a  possible period,
however,  the analysis  made  with standard  algorithms  based on  the
Fourier transforms and  least squares methods provided  no evidence of
that (Fig.\,\ref{F5}).

%

%
\section{Discussion} 
%
%

\subsection{Dust tail}
Comet 2I/Borisov morphology looks similar  to that of comet C/2014\,B1
(Schwartz), with  an equatorial  ejection of  dust particles  from the
nucleus  and   a  near-equatorial   viewing  perspective   (Jewitt  et
al.\,2019b).  However,  since  comet  C/2014\,B1 was  at  about  9\,AU
distance from the  Sun, it showed a faint dust  tail with a brightness
just  above the  sky  background, and  only  visible through  isophote
visualization (See  Fig.\,8, available  only as  supplementary on-line
material).  On  the  contrary,  comet  2I/Borisov  is  at  much  lower
heliocentric distance, shows a greater  emission of gas and dust. Here
solar gravity and solar radiation  pressure are shaping the morphology
of the  dust coma,  and its tail  is in fact  obvious and  oriented in
nearly anti-Solar  direction.  Dust emission in  comet 2I/Borisov seem
to be less collimated than  C/2014\,B1, with the dust particles widely
scattered in anti-Solar direction, possibly  due to the combination of
a lower ejection speed and the effect of a greater radiation pressure.
Our estimate  of the  size of  dust particles  in the  range of  30 to
100\,$\mu$m radius is consistent with that recently reported by Jewitt
(Jewitt et al.\,2019a). However, it  should be considered that in both
cases the  estimates apply to  particles displaced  to the end  of the
visible tail in order to reproduce its apparent length. In fact, using
higher $\beta$ values  in the Finson-Probstein diagram,  in the region
close to the nucleus the  simulation approaches the wide scattering of
the emitted dust  observed in the \textit{HST} images,  which might be
suggestive of the presence also of dust particles with a smaller size.

\subsection{An emissive rotating structure}
Assuming that the nucleus of comet 2I/Borisov is rotating and that the
geometric  position of  the spin  axis is  such to  favor a  prolonged
insolation,  the dust  emission from  a single  source located  on the
sunlit  hemisphere  should be  modulated  by  the insolation  and  the
resulting features of the inner coma should provide indications on the
orientation of the spin axis.  In  the case of two emissive structures
with similar  morphology that  are observed in  approximately opposite
directions, it  could be supposed  that both originated from  the same
active  source  on  a  nucleus  that is  rotating  with  a  spin  axis
coinciding with the axis of the  angle between the two structures and,
if so, that they would be more evident when their position corresponds
to the projection on the plane of the sky for the greater condensation
of material on the perpendicular to the view (Sekanina, 1987).

The finding of  a jet at PA  180$^\circ$ and of its  counterpart at PA
10$^\circ$ suggest that the active  area originating the structures is
located  in  a  near-equatorial  position, and  that  the  nucleus  is
probably rotating with a spin axis lying near the plane of the sky and
geometrically    oriented    at    PA    $\sim$100$^\circ$    or    PA
$\sim100^\circ+180^\circ$  (Fig.\,\ref{F3}).  Unfortunately, from  the
\textit{HST} images it  is not possible to determine  the direction of
rotation of  the nucleus.  To  further verify  our findings, we  run a
computer  model  of  the  inner  coma  using  a  proprietary  software
(P. Pellissier,  2009) specifically designed to  reproduce Earth-based
observations of the dust coma structures and of the development of the
tail. The parameters needed to run the model are the properties of the
dust particles, those of the  nucleus, and the geometric conditions of
the Sun-Earth-comet  system. The following physical  parameters of the
dust particles, compatible  with the observations of  Rosetta on comet
67P (Fulle, 2015;  Rotundi, 2015; Guettler, 2019) were  entered in the
model   (\textit{albedo:   0.03;   diameter:   50   $\mu$m;   density:
  0.6g/cm$^{-3}$;   ejection  velocity:   25  ms$^{-1}$;   dispersion:
  0.75}). For  the cometary nucleus  (assumed spherical) we  entered a
latitude of $5^\circ$ for a single jet  at and a direction of the spin
axis a PA $100^\circ$. Since the rotation period was not known, we run
the model applying only three full  rotations in order to simulate the
early  development of  the inner  coma features  before they  would be
hidden by  the dust. The  resulting model is  shown in Figure  4b. The
features shown by the model are  consistent with those observed in the
\textit{HST} image processed with  spatial filters. By applying higher
number of rotations to the nucleus,  it was also possible to model the
development of  the tail, which  appeared similar to that  observed on
the \textit{HST} image.

\begin{figure}
\begin{center}
\includegraphics[width=80mm]{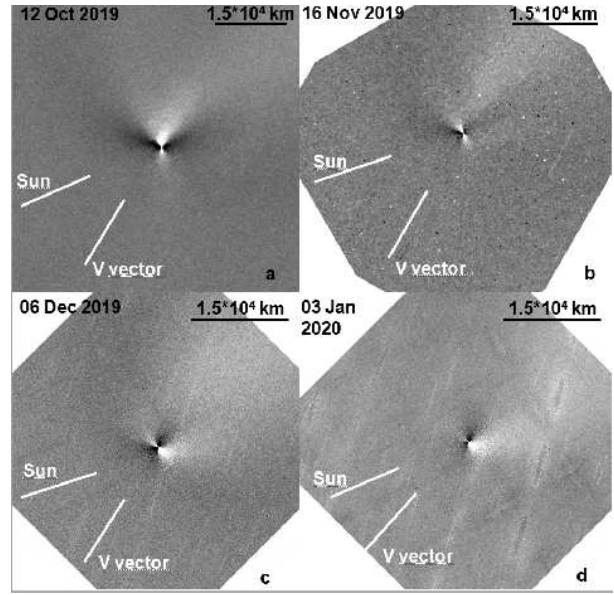}
\caption{
The  \textit{HST}  images taken  in  four  different dates  have  been
processed with  the same  procedure. A L-S  filter ($\alpha=45^\circ$)
has been applied  to highlight micro-contrasts in the  inner coma. The
jet visible in  PA $\sim180^\circ$ on the first date  becomes more and
more  evident   over  time,  while   the  opposite  happens   for  the
contralateral  structure in  PA $\sim10^\circ$.  The direction  of the
spin  axis (assumed  as  the bisector  of the  angle  between the  two
structures) changes due  to the variation of  the geometric conditions
of observation from Earth.
\label{F6}
}
\end{center}
\end{figure}

Our   findings  gathered   stronger  evidence   by  analysis   of  new
\textit{HST}  images taken  in  the subsequent  dates,  thanks to  the
changes   in    the   geometric   conditions   of    the   observation
(Fig.\,\ref{F6}).  In  these images  the  jet  that appeared  directed
towards PA  $\sim180^\circ$ on 12  October 2019 becomes more  and more
evident  and  subsequently  takes  the  shape of  a  fan  due  to  the
variations   of  insolation.   These   findings   also  confirms   the
near-equatorial position of the emitting  source and thus the rotation
of the nucleus.

The availability  of \textit{HST} observations extended  over a period
of about  80 days  has also  allowed us  to study  the changes  in the
direction of  the axis  of the  angle between  the two  observed jets,
which remained lying on the plane of the sky, as confirmed by the fact
that  the  jet  never  showed  evolution  into  shells  or  bow-shaped
structures (Fig.\,9, animation available only as supplementary on-line
material). Assuming  that the  nucleus of the  comet is  spherical and
that the  active area is  located within $10^\circ$ latitude  from the
equator, by means of a trial and  error analysis we found the best fit
for the direction of the spin  axis at RA = 17h20m $\pm$15$^\circ$ and
Dec = $-$35$^\circ$ $\pm$10$^\circ$ (or in opposite direction).

The time span of only $\sim$7\,h is too short to allow for a detection
of a possible  periodicity of the rotation of the  cometary nucleus by
means of a photometric analysis. In fact, no significant variations in
the  light curve  were  detected  also by  Bolin  et  al. (2019),  who
hypothesized that the coma of 2I/Borisov may be too compact to see the
rotational variation,  or that  the nucleus  itself is  spherical, has
slow   rotation   period   or   is  oriented   with   an   unfavorable
geometry. Despite  the intrinsic  limitations of our  measurements, we
think  that  the  small  variations   that  we  could  detect  in  the
instrumental  magnitude  by  applying  a smaller  aperture  suggest  a
confirmation  that  the  nucleus  is rotating,  although  it  was  not
possible to estimate  any period with the available data.  A series of
high-resolution  images over  a longer  time period  could detect  the
period with greater accuracy.
\section{Conclusions}
%
From  our   analyses  of  the   recent  \textit{HST}  images   of  the
interstellar comet 2I/Borisov we found that:
\begin{itemize}

\item The comet  is strongly emitting dust, and producing  a tail that
  is approximately aligned with the anti-Solar vector with a sky-plane
  length $L=8.1\times10^{4}$\,km.

\item Modeling  of the tail suggests  that the dust grains  have radii
  between  30 and  100 $\mu$m,  although  the wide  scattering of  the
  emitted dust  might also  be suggestive of  the presence  of smaller
  dust particles around the nucleus.

\item A clear  elongation of the peri-nuclear area  appears related to
  the presence of  a jet probably deriving from an  active source in a
  near-equatorial position. At least part of  the dust coma and of the
  tail most likely originate from this jet.

\item The nucleus is probably rotating with a spin axis lying near the
  plane of the sky and geometrically oriented at PA $\sim100^\circ$ or
  PA  $\sim100^\circ  +  180^\circ$  on 12  October  2019.  The  small
  variations  detected in  the instrumental  magnitude values  confirm
  this finding, although  they do not allow to  estimate a periodicity
  of the rotation of the nucleus.
 
\item{The  spin  axis   direction  was  determined  at   RA  =  17h20m
  $\pm$15$^\circ$ and Dec = $-$35$^\circ\pm10^\circ$.}

\end{itemize}
In summary, the preliminary findings  of our analyses suggest that the
interstellar  comet  2I/Borisov  appears  similar  in  morphology  and
behavior to the native Solar System comets.
%
%
\section*{Acknowledgments}
This  research has  made use  of the  Keck Observatory  Archive (KOA),
which is operated by the W. M. Keck Observatory and the NASA Exoplanet
Science Institute (NExScI), under contract with the NASA.

%

%


\label{lastpage}


\end{document}